\documentstyle[12pt]{article}
\newtheorem{theorem}{Theorem}
\newtheorem{lemma}{Lemma}
\title{Generalized quantization scheme for central extensions
       of Lie algebras

\author{A A Balinsky\thanks{E-mail: \ mar4751@technion.bitnet} \\
 \small Technion-Israel Institute of Technology \\
 \small Department of Mathematics \\
 \small 32000 Haifa, Israel }

\date{}

\begin{document}

\maketitle

\begin{abstract}  \small
  We present the method for finding of the nonlinear
  Poisson-Lie groups structures on the vector spaces and for their
  quantization. For arbitrary central extension of Lie algebra
  explicit formulas of quantization are proposed.
\end{abstract}

{\bf 1.} Well know that a structure of Poisson-Lie group          
on a vector space ( considered as a commutative group
according to addition ) is defined by an arbitrary
Poisson bracket  depending linearly on the coordinate
functions, i.e.  by  a Lie algebra structure on an adjoint
space. Quantization of  such Poisson bracket lead to
the universal enveloping algebra of this Lie algebra
\cite{R}, and the commutativity of coproduct
follows from the commutativity of  vector space.
Therefore we can to attempt to deformate of  group structure
on the vector space and in such way to obtain the noncommutative
and noncocommutative Hopf algebras \cite{D,FRT}. We have the following
model example.    \\

 {\it Example}.  Consider on $R^{3}$  the following  Poisson bracket
\[ \{ H,x \}=x \ , \ \{ H,y \}=-y \ , \ \{ x,y \}=\frac{\sinh \gamma
H}{\gamma},
\ \ \gamma \neq  0 \  \ \mbox{is parameter} . \]

This bracket has the  properties:
\begin{itemize}
   \item The bracket  compatible with the following coproduct on
   $C^{\infty}(R^{3})$ \\
   $\Delta (x)=x \otimes \exp{(\gamma H/2)}+ \exp{( - \gamma H/2) }
   \otimes x$, \\
   $\Delta (y)=y \otimes \exp{(\gamma H/2)}+ \exp{( - \gamma
   H/2)}
   \otimes y $,  \ \
   $\Delta (H)= H \otimes 1 + 1 \otimes H$, \\
  which can be obtained from the group operation on $R^{3}$:

\[ \left( \begin{array}{c} h \\ x \\ y \end{array} \right) \ast
   \left( \begin{array}{c} h' \\ x' \\ y' \end{array} \right) =
   \left( \begin{array}{c} h+h' \\ x \exp{(\gamma h' /2)} + \exp{( - \gamma h
/2
    y \exp{(\gamma h' /2)} + \exp{( - \gamma h /2)} y'   \end{array} \right)
\]
  Hence,  $(R^{3}, \ast )$ is the Poisson-Lie group
  \cite{D,WL}.
  \item The simple substitution of this bracket by the commutator leads to
the quantum $sl_{q}(2)$.
\end{itemize}

In this paper we want to study the  method for finding of  such
  Poisson-Lie groups structures on the vector spaces and  their
  quantization.

{\bf 2.} The simple way to obtain a structure noncommutative
Lie group on a vector space is the consideration of  commutative
extension of  commutative group. Let  {\bf H}  and {\bf V}    vector
spaces,    and  $\rho_{1} $ and $\rho_{2}$   two commuting
 representations of {\bf H} ( as  the
commutative Lie group ) on the space {\bf V}    . Then we can introduce the
following Lie group $\stackrel{\sim}{G}=$({\bf H$\oplus$V},$\ast$) , where

\begin{equation}
(h,v) \ast (h',v') = (h+h', \rho_{1}(h)v' + \rho_{2}(h')^{-1} v)
                                                                     \label{gr}
\end{equation}

For this group the point (0,0) is the unit and
 \[ (h,v)^{-1}=(-h,- \rho_{1} (-h) \rho_{2} (h) v). \]
The coproduct $\Delta$ , counit $\epsilon$ and the antipode $S$
on  $C^{\infty}( \stackrel{\sim}{G})$
 of the linear functions have the same form
that the corresponding maps in \cite{LM}.
In \cite{LM}  (cf. \cite{GR}) was proposed to use
 such coproduct for definition of  the
deformation of the Lie algebra structure $\Im$    on the space
({\bf H$\oplus$V}$)^{\ast}$      , i.e.
let $\Im$ is the Lie algebra, generated by $H^{i}$ and $X^{m}$
where $H^{i}$ form
the basis of an Abelian subalgebra. Write down the deformation of the
product in $U(\Im )$     , which preserves ($\Delta,\epsilon, S $) :

\[   [x^{r},x^{s}]=[x^{r},x^{s}]_{0} + \Phi ^{rs}(H^{k}; \rho_{1},
   \rho_{2})    \]

Here $[x^{r},x^{s}]_{0}$  is the initial composition
and the deforming functions  $\Phi ^{rs}(H^{k}; \rho_{1}, \rho_{2})  $
are depended on $ \rho_{1}, \rho_{2}$ and are the power series of
 $H^{i}$ . The question of the
possibility of  such deformation leads to the question of the possibility
to define  a  Poisson-Lie structure  on the group  $\stackrel{\sim}{G}$
   for which global
Poisson bracket of the functions $\{ h^{i} \} $  is equal to zero and Poisson
bracket of the coordinate functions  differs from the linear one only on
the functions of  $\{ h^{i} \} $.       It will be good for the
quantization of such system, because we have not the problem with ordering.
Therefore we have the problem of the finding globally Poisson-Lie
bracket on the group  $\stackrel{\sim}{G}$
 from the cocycle  $\delta : g \rightarrow g \otimes g $,
where  $g$ is the Lie algebra of $\stackrel{\sim}{G}$
and $\delta^{\ast}$  is our \ $[.,.]_{0}$ \ structure of
Lie algebra on the space ({\bf H$\oplus$V}$)^{\ast}$ with
$[h^{i},h^{j}]=0$ for all $i,j$.
The group  $\stackrel{\sim}{G}$    topologically is
 vector space and hence such Poisson bracket
exists globally. The general analysis when such bracket   is linear + function
of $\{ h^{i} \} $ will be done in a forthcoming paper.  In
general case we have

\begin{lemma} If we have the cocycle $\delta$ for which
{\bf $H^{\ast}$ } is an Abelian subalgebra,
 the global Poisson-Lie bracket
 of the functions $\{ h^{i} \} $ on $\stackrel{\sim}{G}$ is equal
 to
 zero.
\end{lemma}

Now we are going to give the exposition of the method for finding
 of global Poisson bracket on   $\stackrel{\sim}{G}$  from cocycle $\delta$  .
After the change of variables $(h,v) \mapsto (h, \rho_{2} (h) v)$
we obtain the group structure on $\stackrel{\sim}{G}$
 without $\rho_{2}$:
\begin{equation}
(h,v) \bullet (h',v') = (h+h', \rho(h)v' + v)           \label{grn}
\end{equation}
wherein $\rho (h)= \rho_{1} (h)   \rho_{2} (h) $.
 Note that the differential of the change in the
 unit of $\stackrel{\sim}{G}$  is {\bf Id}  , i.e. this
change of variables does not change  of Lie bialgebra structure
on  $g$. Further we work with  ($\stackrel{\sim}{G}$,$\bullet$).
 Let $h_{\bullet}$     is infinitesimal
version of $\rho$, i.e. $\rho (h)= \exp{(h_{\bullet})}$.

\begin{lemma}
For the Lie algebra $g$ of the group  (\ref{grn}) ,
  $g=H \oplus V$ with                       \\
  $[h,h']=0, \ [v,v']=0, \ [h,v]=h_{\bullet}v$.
\end{lemma}

For the Lie group $\stackrel{\sim}{G}$ the constant vector field
$(h',0)$ is left-invariant for all $h' \in$H
  and the constant vector field $(0,v')$
       is right-invariant
for all $v \in$V. If $\pi^{ij}(h,v)$ is the Poisson-Lie bracket  of the
coordina
functions then from the Theorem 1.2 in \cite{WL}  we have

\begin{lemma} For all $k$ and $l$    tensor
$\partial \pi / \partial h^{k}$ is a left-invariant bivector field
and
$\partial \pi / \partial v^{l}$ is a right-invariant bivector
field on the group  $\stackrel{\sim}{G}$.
\end{lemma}

But $\left.(\partial \pi^{ij} / \partial h^{k}, \partial \pi^{ij} /
\partial v^{l}) \right| _{(0,0)}$ \    is our cocycle $\delta$.
{}From the Lemma 3 and the
property that  $\pi (0,0)=0$         we can find globally  $\pi$      .

{\it Example}. H and V are one-dimensions, $h_{1}.
v_{1}=v_{1}$.
Then   $\stackrel{\sim}{G}$  is the   Poisson-Lie group with bracket
 $\{ h^{1},v^{1} \} = \alpha  v^{1} + \beta (e^{h^{1}} - 1)$, \
 $\alpha , \beta$ - parameters.
The simple substitution of this bracket by the commutator leads to
the quantum   Lie algebra $U_{\alpha, \beta}$    generated by
$h$ and $v$       with
\[      [ h,v ] = \alpha  v + \beta (e^{h} - 1),   \]
\[     \Delta (h)= 1 \otimes h + h \otimes 1, \ \Delta(v)=e^{h}
     \otimes v + v \otimes 1 ,  \]
\[     S(h)=-h, \ S(v)=e^{-h} v,\ S(1)=1,\ \epsilon (h)=
     \epsilon (v)=0, \epsilon(1)=1. \]

{\bf 3.}  Consider an arbitrary central extension of Lie algebra.
Let {\bf V}=$\langle v^{i} \rangle$ algebra Lie and $\langle
h^{1} \rangle  \oplus${\bf V} its
central-extension with cocycle $\Omega^{ij}_{0}$, i.e.
\begin{equation}
[h^{1},v]=0 \ \ \ \forall v \in V, \ \ [v^{i},v^{j}]=
\Omega^{ij}_{0} h^{1}
+ C^{ij}_{k} v^{k},                  \label{ext}
\end{equation}
where $C^{ij}_{k}$ are the structuring constants of the Lie algebra {\bf
V} (here and below we assumed a summation on repeat indexes).
Let $\alpha$ and $\beta$ are two commuting differentiations of {\bf V}. Then
$\stackrel{\sim}{G}$= $\langle h_{1} \rangle \oplus${\bf V}$^\ast$
is the Poisson-Lie group with the product (\ref{gr}) , where
$\rho_{1} (h_{1})=e^{\alpha^{\ast}}$, $\rho_{2}
(h_{1})=e^{\beta^{\ast}}$,
 and with the cocycle adjoint to  (\ref{ext}).
On $\stackrel{\sim}{G}$
 coproduct, counit and antipode  have the following form

\[ \Delta (h^{1}) = h^{1} \otimes 1 + 1 \otimes h^{1},  \]
\begin{equation}
\Delta (v^{i}) = e^{h^{1} \otimes \alpha}(1 \otimes v^{i})
+  e^{- \beta \otimes h^{1}}( v^{i} \otimes 1),       \label{copr}
\end{equation}
\[  S(h^{1})= - h^{1},\ S(v^{i})=- exp (-h^{1} \alpha ) exp
(h^{1} \beta ) v^{i} , \ \epsilon (h^{1}) = \epsilon (v^{i})=0  \]

\begin{theorem}
The global Poisson-Lie bracket on  $\stackrel{\sim}{G}$
with the cocycle adjoint to (\ref{ext}) equals:
\begin{equation}
\{ h^{1},v^{j} \}=0 \ \ \ \forall j, \ \ \ \{v^{i},v^{j} \} =
\Omega^{ij}(h^{1}) + C^{ij}_{k} v^{k} ,     \label{osn}
\end{equation}
where  \\
        \\

$ \left| \left| \Omega^{ij}(h^{1}) \right| \right| =   $
\[       = exp(-h^{1} \beta) \cdot \left( \int_{0}^{h^{1}}
        exp (h'(\alpha + \beta)) \cdot \| \Omega^{ij}_{0} \|
        (exp (h'(\alpha + \beta)))^{t}dh' \right) \cdot
        (exp(-h^{1} \beta))^{t}  \]  \\

Here t is transposition of matrix and
$\left| \left| \Omega^{ij}  \right| \right|$ is a matrix
with the entries of $\Omega^{ij}$.
\end{theorem}

\begin{theorem}
 The simple substitution of this bracket  (\ref{osn})   by the commutator leads
the quantization of the algebra (\ref{ext}), compatible with
(\ref{copr}).
\end{theorem}
 From these Theorems we have the large class
 of the quantum groups generated by   $h^{1}$  and $v^{i}$
\  with the relations (\ref{osn})
and   ($\Delta , S   , \epsilon $)   from (\ref{copr}).
Example 1 from \cite{LM} is a special case when {\bf V}   is a
 two-dimensional commutative
Lie algebra.

In a forthcoming paper we shall describe in details the case of the
Virasoro algebra.

\pagebreak


\begin{thebibliography}{99}                                        

\bibitem{D}
 Drinfeld V G  1986 Quantum groups
\newblock {\it Proc. of the International Congress of Math.},
           {\bf vol 1} (Berkeley, CA: Academic) pp 798-820

\bibitem{LM}
 Lyakhovsky V and  Mudrov A 1992
\newblock {\it J. Phys. A:Math. Gen.} {\bf 25}, L1139-L1143

\bibitem{WL}
 Weinstein A and Lu J 1990
\newblock {\it J. Diff. Geometry}, {\bf 31} 501-526

\bibitem{R}
Rieffel M A 1988
\newblock {\it ``Lie group convol. algebra as deform. of linear poisson
                 struct.''}, Math. preprint

\bibitem{FRT}
 Faddeev L D, Reshetikhin N Yu and Takhtajan L A 1989
\newblock {\it Alg. Anal}, {\bf 1} , 178

\bibitem{GR}
 Grossman R and  Radford D 1992
\newblock {\it Contemporary Mathematics}, {\bf 134}, 115-117

\end{thebibliography}
\end{document}